\documentclass[final,5p,times,twocolumn]{elsarticle}

\usepackage{graphicx}
\usepackage{amssymb}
\usepackage{amsmath}
\usepackage{enumitem}
\usepackage{multirow}
\usepackage{tikz}
\usepackage{epstopdf}
\usepackage[percent]{overpic}
\usepackage{scrextend}
\usepackage{xcolor,xspace}
\usepackage[ulem=normalem]{changes}
\usepackage{hyperref}
\hypersetup{colorlinks=True,urlcolor=blue,linkcolor=blue,citecolor=blue,filecolor=black}
\usepackage{dcolumn,color,footnote,bm,braket}
\usepackage{url,longtable,tabularx}
\usepackage{threeparttable}
\usepackage{fancybox}

\usetikzlibrary{matrix}

\colorlet{Changes@Color}{red}

\newcommand\+{\dagger}
\newcommand{\be}[4]{B(E2; #1^+_{#2} \!\to\! #3^+_{#4})}

\journal{Physics Letters B}

\begin{document}

\begin{frontmatter}

\title{
Microscopic derivation of the
interacting boson model parameters
with machine learning}

\author[1]{Y.~Obata}
\author[1,2]{K.~Nomura\corref{cor}}
\ead{nomura@sci.hokudai.ac.jp}

\affiliation[1]
{
organization={Department of Physics, Hokkaido University},
city={Sapporo},
postcode={060-0810},
country={Japan}
}

\affiliation[2]
{
organization={Nuclear Reaction Data Center, Hokkaido University},
city={Sapporo},
postcode={060-0810},
country={Japan}
}

\cortext[cor]{Corresponding author}

\begin{abstract}
Machine learning is applied to
derive microscopically parameters
of the interacting boson model
for nuclear spectroscopy.
A physics-guided neural network is proposed,
which is trained to map the
potential energy landscapes
that are calculated
within the nuclear density functional theory
onto the bosonic parameter space.
To incorporate the underlying
nuclear structure information and mitigate
parameter degeneracy,
the network integrates a global
quadrupole collectivity indicator
and valence nucleon
numbers as key input features.
In its applications to rare-earth nuclei,
by reproducing
the microscopic energy
landscapes without any manual parameter
tuning,
the trained network is shown to provide
a set of the model parameters and
energy spectra that
reflect the nuclear structural evolution,
offering a robust
alternative microscopic description
of nuclear collectivity.
\end{abstract}

\date{\today}


\end{frontmatter}

The atomic nucleus is a quantum
many-body system that exhibits a
rich variety of collective phenomena
\cite{BM,RS}, prominent examples being
the quantum phase transition (QPT)
\cite{cejnar2010,iachello2011a,fortunato2021}
and coexistence
\cite{wood1992,heyde2011,garrett2022,leoni2024}
of ground-state shapes.
The microscopic description
of these phenomena
based upon the underlying nucleonic
degrees of freedom has been
a central theme in nuclear physics.
The interacting boson model (IBM)
\cite{IBM} has been successful in studying
low-energy quadrupole collective states
in medium-heavy and heavy nuclei.
The model comprises
$s$ and $d$ bosons, which reflect
correlated monopole
and quadrupole pairs of valence
nucleons, respectively
\cite{OAIT,OAI}.
Conventionally the IBM studies
have relied on the phenomenological
fit of the parameters to the
experimental low-energy spectra,
hence attempts have been made to
derive the IBM parameters
from microscopic nuclear structure
models, e.g.,
the nuclear shell model \cite{OAIT,OAI},
in limited realistic applications to
nearly spherical nuclei
\cite{otsuka2020microscopic,mizusaki1997}.

A microscopic formulation of the
IBM in the general
cases of the quadrupole collective states
was developed in
Refs.~\cite{nomura2008,nomura2010,
nomura2011rot,nomura2012tri},
in which the potential
energy surface (PES) calculated
self consistently within the
nuclear density functional theory (DFT)
\cite{RS,bender2003,vretenar2005,robledo2019}
is mapped onto the equivalent
energy surface in the boson system.
This procedure specifies the IBM Hamiltonian
without any adjustment
to experimental data.
The method bridges the gap between
the DFT and IBM, and has allowed for
studying various nuclear
structure phenomena
(see Ref.~\cite{nomura2025rev}
and references therein).
In this scheme,
the strength parameters of the IBM
Hamiltonian are determined
so as to reproduce basic characteristics
of the DFT PES,
such as the location of and the
curvature around the energy minimum.
However, since this optimization
is performed for each nucleus,
careful adjustments are often required
to ensure smooth systematic behaviors
of the parameters along isotopic chains.

In this work, we address
the above limitation of the
PES-mapping by means of
machine learning (ML).
Nowadays ML has been
employed in vast domains of physics,
including nuclear physics
(see Ref.~\cite{boehnlein2022}
for a recent review).
In the nuclear DFT,
ML was adopted to emulate PESs
for the studies of shape and
deformation properties
\cite{akkoyun2013,lasseri2020,lay2024}.
Of particular interest here is
the application of the ML technique
to analyze spectroscopic
properties, that is,
collective excitations
and shape phase transitions,
and the relevant applications were
made recently in a number of
empirical studies based on the
experimental data,
e.g., Refs.~\cite{lv2025,Ahmad2026,zhang2026,li2026}.
Specifically, in Ref.~\cite{Ahmad2026}
Light Gradient Boosting Machine
algorithm \cite{lightGBM} was combined
with the IBM to study shape phase transitions
in rare-earth nuclei.
Such pure ML approaches
were designed to reproduce
experimental data phenomenologically,
but they have often been used as a black box
and thus the connection
to the underlying microscopic nuclear
structure is not clear.
Therefore, a hybrid approach that combines
the efficiency of ML with
a certain microscopic theory is required.

The aim of this work is to bridge this gap
by establishing a microscopic derivation
of the IBM Hamiltonian by means of ML.
For this purpose, we propose
a physics-guided neural network (PGNN).
Rather than the conventional iterative
fitting, we train a neural network (NN)
to map the DFT PESs directly to the
IBM parameter space.
A key feature of our approach is
the integration of a differentiable
physics layer within
the network architecture \cite{Raissi2019}.
This layer explicitly calculates
the bosonic PES from the predicted
parameters during the training process,
ensuring that the network optimization
is strictly guided by the
underlying microscopy of the IBM.
We apply this method to
Nd, Sm, and Gd isotopes with
the neutron number $N=86$ to 100,
which are known to exhibit the QPTs
from nearly spherical
to strongly deformed shapes at $N\approx90$
\cite{iachello2001X5,casten2001X5,cejnar2010},
characterized by the abrupt change
of observables such as
the ratio $R_{4/2}\equiv E(4^+_1)/E(2^+_1)$.
We then demonstrate that the PGNN
can systematically extract Hamiltonian
parameters with high accuracy and
without manual tuning;
the derived parameters exhibit
nucleon-number dependence that
reflects the underlying nuclear
structure, and reproduce
the nuclear structural evolution
consistently with experiment.

The microscopic description
of the nuclear shapes is here based on
the Hartree-Fock-Bogoliubov
(HFB) theory \cite{robledo2019}
employing the D1S set \cite{D1S}
of the Gogny force \cite{Gogny},
which allows for robust and systematic
descriptions of nuclear structure
across the entire regions of
the nuclear chart
\cite{robledo2019}.
The constrained Gogny-D1S HFB
calculations provide
energy surfaces as functions
of the axial quadrupole
deformation, $\beta$, and
triaxiality, $\gamma$.
As an initial study,
we restrict our analysis
to axially symmetric configurations,
namely, the $\gamma$
degree of freedom is not taken into account.
Thus the one-dimensional HFB energy surfaces,
denoted potential energy curves (PECs),
which are functions of $\beta$ only,
shall be used as training data
for our NN.
These HFB PECs are adopted from the
AMEDEE database \cite{hilaire2007,CEA}.
The deformation $\beta$ is related to
the mass axial quadrupole moment,
$\hat{Q}_{20}$, used as a constraint
in the HFB calculations, and is expressed
as \cite{BM,CEA}
$\beta = {\sqrt{5\pi}}/(3A R^2) \braket{\hat{Q}_{20}}$,
where $R=1.2A^{1/3}$ (in fm)
and $\braket{\hat{Q}_{20}}$ denotes
the expectation
value of the $\hat Q_{20}$
in the HFB ground state.

In this study we discuss the
proton-neutron IBM
(denoted IBM-2 hereafter), which distinguishes
like-neutron bosons from like-proton
bosons.
In the IBM-2,
the numbers of proton and neutron bosons,
denoted $n_\pi$ and $n_{\nu}$,
are equal to those of valence
proton and neutron pairs, respectively.
The total number of bosons,
and the numbers of
proton and neutron bosons separately
are conserved for a given nucleus.
The Hamiltonian of the IBM-2 adopted
here is of the form,
\begin{equation}
\hat{H}_{\text{IBM}}
= \epsilon (\hat{n}_{d_\pi} + \hat{n}_{d_\nu})
+ \kappa \hat{Q}_\pi \cdot \hat{Q}_\nu
\; .
\label{eq:hb}
\end{equation}
The first term represents the
number operator for $d$ bosons,
given by
$\hat{n}_{d_\rho} = d^\dagger_\rho \cdot \tilde{d}_\rho$
($\rho=\pi,\nu$),
with $\epsilon$ being the
single-$d$ boson energy with respect
to that of $s$ bosons.
The second term stands for the
quadrupole-quadrupole interaction
between proton and neutron bosons
with the strength $\kappa$,
and $\hat{Q}_\rho$ is the bosonic
quadrupole operator defined as
$\hat{Q}_\rho =
s^\dagger_\rho \tilde{d}_\rho + d^\dagger_\rho \tilde{s}_\rho
+ \chi_\rho [d^\dagger_\rho \tilde{d}_\rho]^{(2)}$,
where $\chi_\pi$ ($\chi_{\nu}$)
is a dimensionless
parameter that determines if the intrinsic
shape of the proton (neutron) fluid
is prolate or oblate deformed,
according to whether it is negative
or positive in sign, respectively.
The Hamiltonian \eqref{eq:hb}
was frequently used in realistic
IBM-2 calculations
\cite{IBM,OAI,nomura2008,nomura2010},
and was shown in microscopic studies
\cite{OAIT,OAI,nomura2008}
to represent essential correlations
that come into play in
low-energy quadrupole collective states.
The corresponding parameters,
$\{\epsilon, \kappa, \chi_\pi, \chi_\nu\}$,
are thus essential for describing
quadrupole shapes and collective
excitations, and we shall derive them
microscopically by the procedure
described below.

The connection between the DFT
and IBM-2 can be established by matching
the PEC of the latter to
that of the former \cite{nomura2008}.
The intrinsic state of the boson system
is defined as the coherent state
\cite{dieperink1980,ginocchio1980a,bohr1980},
which reads
\begin{equation}
\ket{\Phi}
=
\prod_{\rho=\pi,\nu}
\left(s^{\+}_{\rho}+\beta_{\rho}d^{\+}_{\rho0}\right)^{n_{\rho}}\ket{0}
\; ,
\end{equation}
up to the normalization factor.
The amplitudes
$\beta_{\pi}$ and $\beta_{\nu}$
are boson analogs of the
axial proton and neutron deformations,
respectively,
and the boson vacuum, $\ket{0}$,
corresponds to the inert core,
that is, doubly magic nucleus
$^{132}$Sn.
We assume that the proton and neutron
deformations are equal, i.e.,
$\beta_\pi = \beta_\nu \equiv \beta_{\rm B}$,
and that the bosonic deformation, $\beta_{\rm B}$,
is proportional to the
fermionic one \cite{ginocchio1980b,nomura2008},
such that 
$\beta_{\rm B} = C_\beta \beta$,
with $C_{\beta}$
being a constant of proportionality.
The factor $C_{\beta}$
is introduced to account
for the difference
in the model space
and the degrees of freedom between
the microscopic DFT
and the IBM-2: The former is based on
the full nucleonic degrees of freedom,
whereas the latter is built on
a given valence space.
The IBM-2 PEC,
$E_{\text{IBM}}(\beta)$,
is obtained in an analytical form
by taking the expectation
value of the Hamiltonian \eqref{eq:hb}
in the coherent state $\ket{\Phi}$,
and serves as
the physics layer in our NN.

Here we set the deformation scale factor
to be $C_{\beta}=4.0$ for all nuclei,
which is a typical value considered for
rare-earth nuclei within the IBM-2 mapping
\cite{nomura2008,nomura2010}.
This choice is not only for the
sake of consistency with
the earlier studies, but also
turns out to be
optimal for the mapping using
the PGNN.
When a smaller $C_{\beta}$, e.g.,
$C_{\beta}=2.5$, is chosen,
the IBM-2 PEC becomes
spread in the $\beta$ direction,
and to reproduce the absolute
potential depth of the HFB PEC
the derived values of the
strength parameters
$\kappa$ and $\epsilon$ turn out to
be unexpectedly large in magnitude;
the resulting energy
spectra are higher than the
experimental ones by a factor of three,
which is unrealistic.
For a larger $C_{\beta}$, e.g.,
$C_{\beta}\approx4.5$,
the discrepancy between the IBM-2
and HFB PECs becomes significant;
this is because the constraints
of the $sd$-IBM-2 space come to
limit its ability to replicate
the complex energy landscape
in the region of large $\beta$
deformation.
Among tested values,
we find that the one $C_\beta = 4.0$
leads to the most
balanced description of observables.
That is,
it not only captures the characteristic
jump in the $R_{4/2}$
ratio across the $N=90$ boundary but also
maintains physically reasonable
energy scales of excitation spectra.
We note that
an expression for $C_{\beta}$
was given in Ref.~\cite{ginocchio1980b},
that it is proportional to
the ratio of the mass
to valence nucleon numbers;
resulting $C_{\beta}$,
e.g., for $^{148}$Sm and $^{154}$Sm
are approximately 8 and 6,
respectively.
As mentioned above, however,
these values are too large for the
present mapping procedure
using NN to yield remaining Hamiltonian
parameters that are realistic.
Therefore, instead of using
the expression of
Ref.~\cite{ginocchio1980b},
$C_{\beta}$
is here treated rather as a parameter,
as in the previous IBM-2
mapping calculations
\cite{nomura2008,nomura2025rev}.

\begin{figure}[t]
\centering
\includegraphics[width=\linewidth]{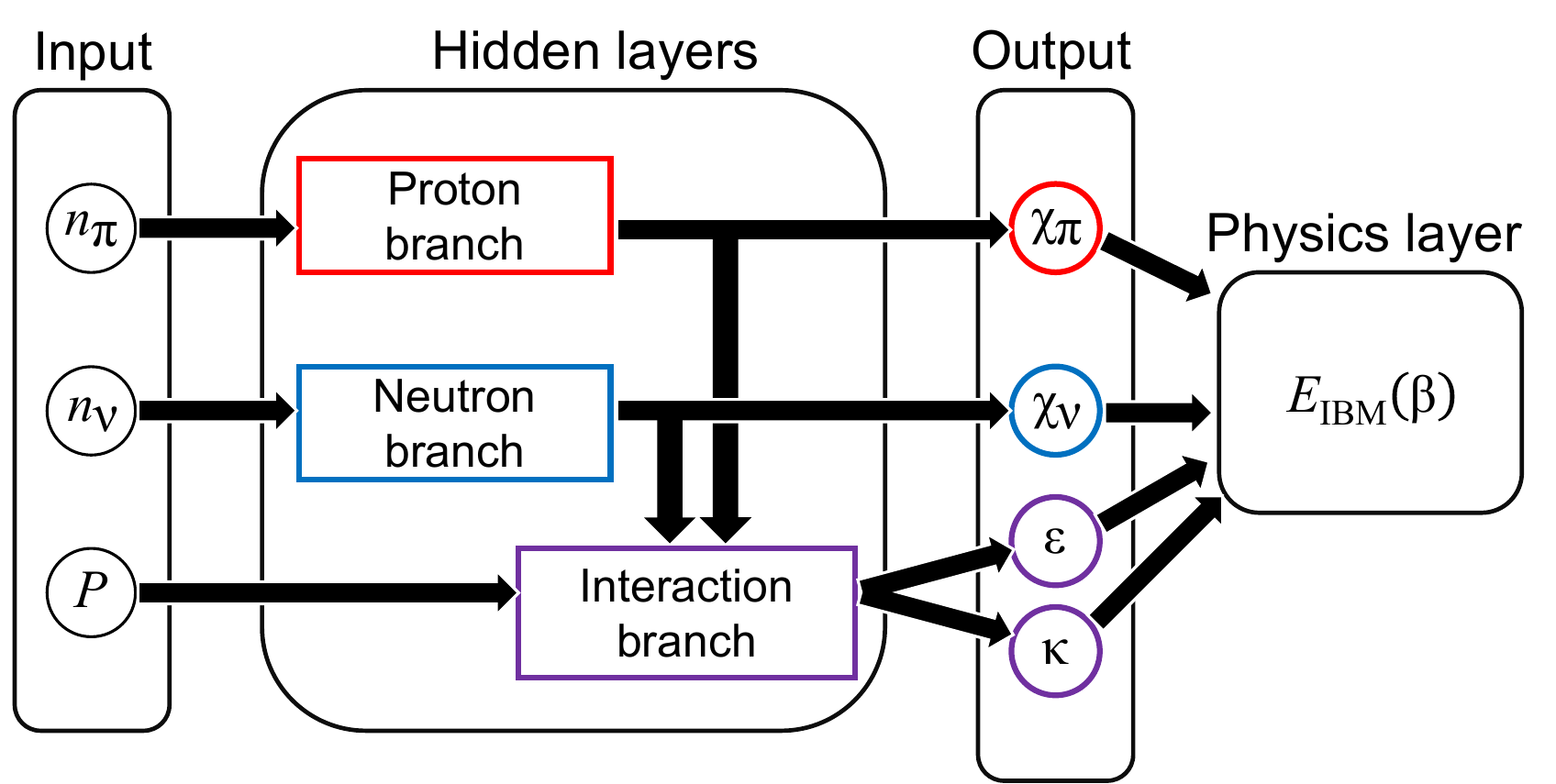}
\caption{Schematic illustration
of the physics-informed branching
architecture.
The network consists of
proton, neutron, and interaction branches
to predict sets of the IBM-2 parameters
$\{\chi_\pi, \chi_\nu, \epsilon, \kappa\}$.
Outputs are fed into a non-trainable
physics layer that computes the IBM-2 PEC,
$E_{\text{IBM}}(\beta)$,
for the loss calculation.}
\label{fig:arch}
\end{figure}

For PGNN,
we develop a physics-informed
branching architecture to extract the essential
empirical features of nuclear structure
evolution across isotopic chains.
To effectively capture the underlying
microscopic structure of the IBM-2,
we design a specific
network architecture that reflects
the distinction between the
proton and neutron boson degrees
of freedom.
A schematic
illustration of the architecture
is given in Fig.~\ref{fig:arch}.
The input vector,
$\mathbf{x}$,
consists of three fundamental quantities:
\begin{equation}
\mathbf{x} = \{n_\pi, n_\nu, P\}
\; .
\end{equation}
Here $P$ is defined as \cite{casten1987}
\begin{equation}
P = \frac{N_p N_n}{N_p + N_n} \; ,
\label{eq:casten_factor}
\end{equation}
with $N_p(=2n_{\pi})$ and $N_n(=2n_{\nu})$
being the numbers of
valence protons and neutrons, respectively.
It represents
the competition between the pairing
and valence proton-neutron interactions, 
and is empirically known
as an indicator
of nuclear deformation and phase transitions
\cite{casten1987,cejnar2010}.
In recent ML analyses,
the $P$-factor was indeed shown to be
most essential for describing
shape phase transitions in rare-earth
nuclei \cite{Ahmad2026},
and for predicting low-energy levels
and $\be 0 1 2 1$ values
in a wide range of the nuclear mass
chart \cite{lv2024}.
By explicitly including the $P$
as an input feature, we aim to
enhance the network's ability to learn
the evolution of deformation
and collectivity.

A major problem for the microscopic
IBM-2 mapping was the appearance of
parameter degeneracy, in which
multiple combinations of
the IBM-2 parameters give rise to
similar PECs.
To mitigate this, we introduce into
the network architecture strong constraints
that the intrinsic-shape parameters,
$\chi_{\pi}$ and $\chi_{\nu}$,
should be determined independently
of $\epsilon$ and $\kappa$,
which govern the competition
between spherical and deformed phases.
In the earlier
microscopic studies of the IBM-2,
such as the Otsuka-Arima-Iachello (OAI) mapping
\cite{OAI}, $\chi_\pi$ and $\chi_{\nu}$
are primarily determined to reflect
the underlying shell structure and
the number of valence nucleons
of given nuclei.
These parameters follow the
systematic expected in the
generalized seniority or
quasi-spin formalism,
and typically change in sign
at the middle of the major shell.
In this study, we assume that the
intrinsic shape of the proton fluid
(inferred from $\chi_\pi$) and that
of the neutron fluid (inferred from $\chi_\nu$),
respectively, depends solely on
$n_\pi$ and $n_\nu$,
respectively.
This assumption also conforms
to the boson-number dependencies
of $\chi_{\pi}$ and $\chi_{\nu}$
found in the earlier
microscopic IBM-2 studies
\cite{scholten1980phd,scholten1983,mizusaki1997}.
To incorporate this physical assumption
as a constraint, we employ
a branched NN architecture
so as to prevent $\chi_\pi$ ($\chi_{\nu}$)
from being influenced by the
inputs of $n_{\nu}$ ($n_{\pi}$) and $P$.

The shape phase transition
is described in the IBM-2
by the interplay between the two terms
in $\hat H_{\rm IBM}$ \eqref{eq:hb},
that is,
the spherical-shape-driving,
pairing correlations, represented by
the $\hat n_d$ term,
and the deformation-driving, proton-neutron
quadrupole interaction, represented by
the $\hat Q_{\pi}\cdot \hat Q_{\nu}$
term.
While the parameters $\chi_{\rho}$
concern the single-species property
(i.e., proton or neutron deformation),
$\epsilon$ and $\kappa$ are treated
as being dependent on both
$n_{\pi}$ and $n_{\nu}$.
As depicted in Fig.~\ref{fig:arch},
the interaction branch introduced
here is designed
to capture these effects,
and aggregates the $P$-factor,
which is a global collectivity indicator,
and the latent feature vectors
extracted from the proton
and neutron branches.
By combining these signals,
the network learns how the specific
shell configurations of protons and neutrons
interfere to determine the subtle
balance between the two terms
of the IBM-2 Hamiltonian,
and effectively predict the
softness and rigidity in
the $\beta$ deformation of the PEC.

To ensure that the predicted Hamiltonian
parameters physically make sense,
we impose strict sign constraints
on the output layer.
Specifically, the single-$d$-boson energy
must be positive ($\epsilon > 0$),
and the quadrupole-quadrupole interaction
must be attractive ($\kappa < 0$);
furthermore,
since rare-earth nuclei are supposed to
be dominated by prolate deformation,
the shape parameters
$\chi_\pi$ and $\chi_\nu$ should be
negative.
These constraints are introduced
by applying the Softplus activation function
(and sign inversion where necessary)
to the network outputs.
The outputs of these branches are
not trained against ground-truth parameters,
which are often ambiguous or unknown.
Instead, they are fed into the physics layer,
a non-trainable differentiable
module implementing $E_{\rm IBM}(\beta)$.
By integrating this physics layer,
the network is trained end-to-end to extract
parameters that simultaneously satisfy
the nuclear structural constraints imposed by the
branching architecture and accurately
reproduce the microscopic deformation
energy landscapes.

To determine the optimal configuration
without manual tuning bias,
we employ the Optuna framework \cite{akiba2019optuna}
for Bayesian hyperparameter optimization.
We perform an automated search over
50 trials to identify the best combination
of the number of hidden layers, neurons per layer,
activation function, learning rate,
and batch size.
The hyperparameter
optimization is focused exclusively
on minimizing the
mean absolute error (MAE)
of the PECs for the validation dataset.
Through the 50 trials of Bayesian optimization,
the network converges to an architecture
that effectively captures the global features
of the microscopic PECs with high precision.

\begin{figure}[t]
\centering
\includegraphics[width=\linewidth]{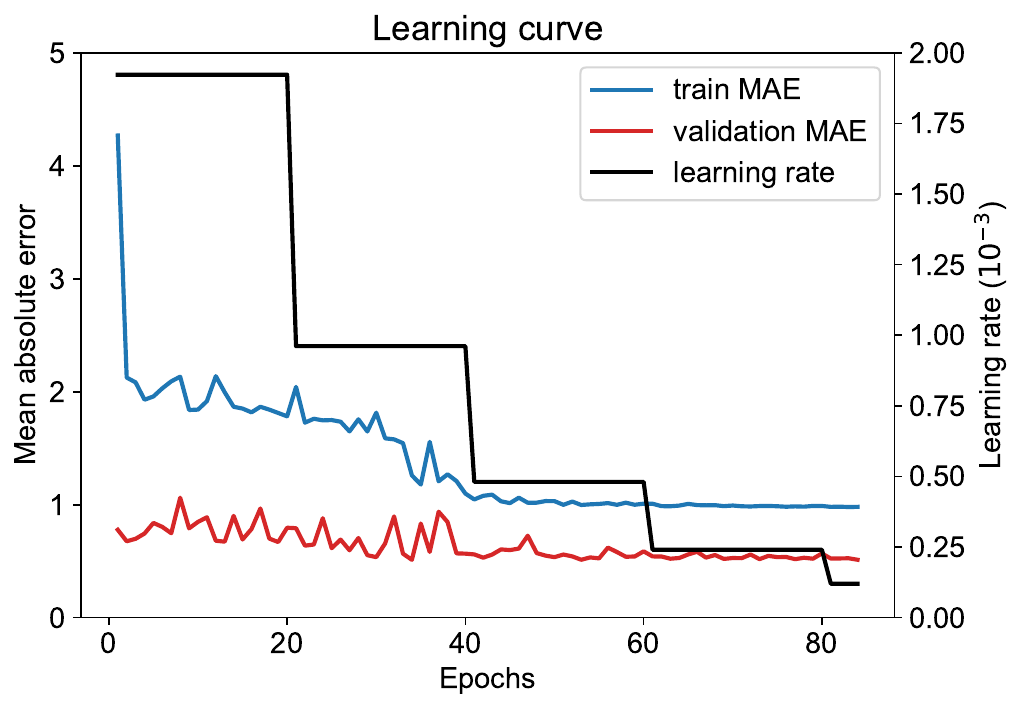}
\caption{
Training and validation loss histories
under optimal hyperparameter configurations
with the $N=90$ nuclei as a validation set.}
\label{fig:loss}
\end{figure}

The training stability of the
NN is analyzed through the learning
curves, shown in Fig.~\ref{fig:loss}.
The model focuses on MAE
to ensure robustness against
the limited number of samples,
i.e., 24 PECs for the
even-even $^{146-160}$Nd,
$^{148-162}$Sm, and $^{150-164}$Gd.
To avoid the risk of selecting
unrepresentative ``endpoints''
(i.e., the nuclei close to $N=86$ or 100)
through random sampling,
we take a specific validation
strategy by designating the isotopes
with $N=90$ from each chain
as the validation set.
As shown in Fig.~\ref{fig:loss},
the validation loss remains
lower than the training loss throughout
the process.
This behavior suggests that the
$N=90$ nuclei, which correspond to
the critical points
\cite{iachello2001X5,casten2001X5}
of the QPTs,
are well described by the global
trend learned from the rest of
the isotopic chains.
Here it is worth a remark that
the number of samples, i.e., 24,
for our NN is small
as compared with those
considered in usual ML-based approaches
to nuclear structure,
and that the choice of the validation set
to be specific nuclei at $N=90$
is made {\it a posteriori} and
is yet to be justified.
The scope of the
present study is, however, to give
a first successful case of the NN
to find optimal IBM-2 parameters even
with the limited numbers of
training and validation sets.
To strengthen the robustness
of our method, the stability of the
results should
be probed with much larger numbers
of training and validation sets.
This, however, remains an open
problem for the present form of
the NN architecture for the
IBM-2 mapping, and
a complementary
NN-architecture model
based on a larger number
of dataset could be developed.

\begin{figure}[t]
\includegraphics[width=\linewidth]{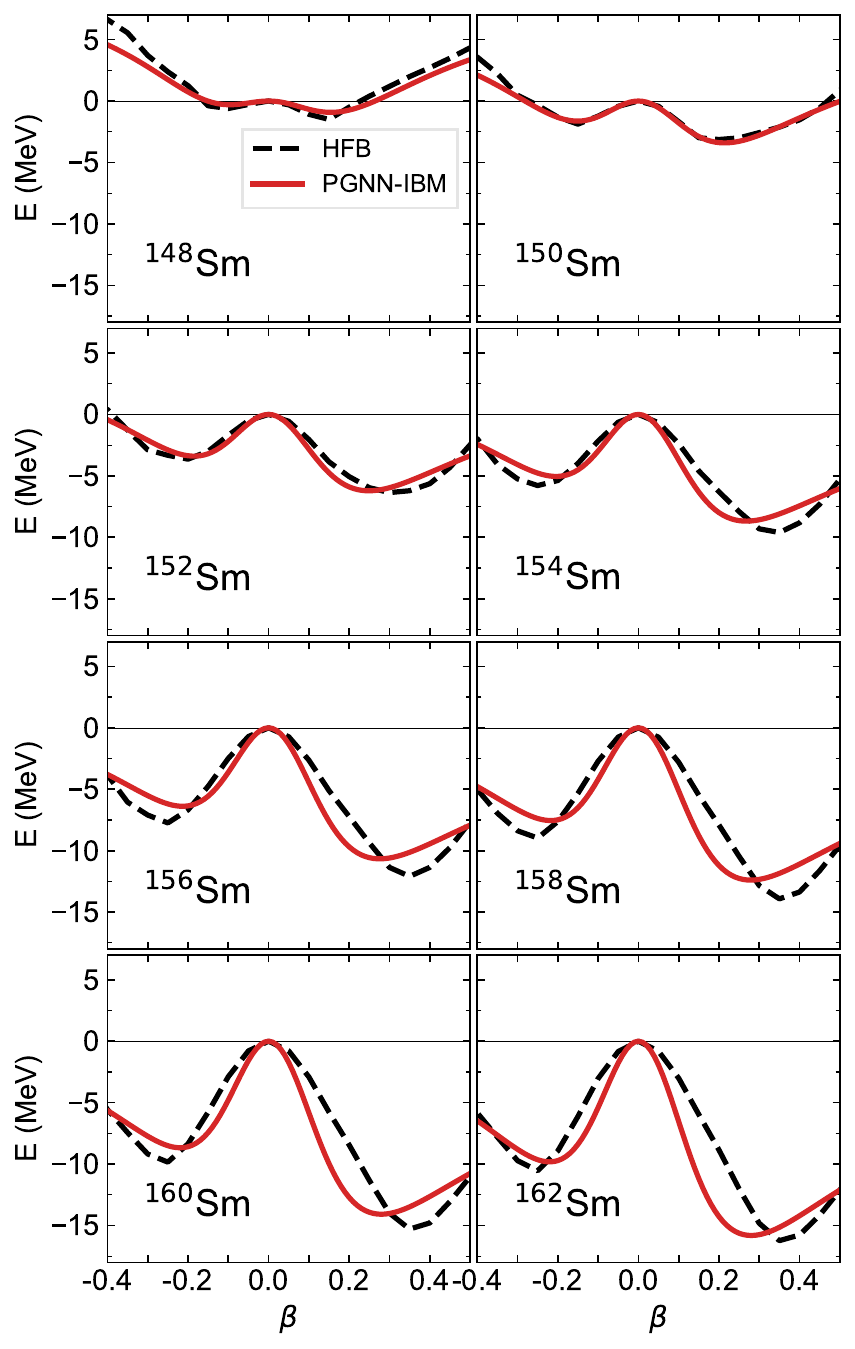}
\caption{Potential energy curves (PECs)
along the axial quadrupole deformation $\beta$
for the even-even $^{148-162}$Sm
isotopes calculated with the Gogny-D1S
HEB \cite{CEA} (dashed curves)
and PGNN-IBM (solid curves).}
\label{fig:pec}
\end{figure}

We now turn to discuss
the performance of the
PGNN in its application
to realistic nuclei,
by taking as an illustrative example
the Sm isotopes.
The conclusions on the quality
of the PGNN-IBM
in describing physical observables
are similar for the Nd and Gd isotopes.
Our analysis focuses on how the
reproduction of microscopic PECs
leads to a consistent description
of nuclear structural evolution. 
The primary criterion to judge
the quality of the PGNN
is, therefore, whether it is
able to map accurately
the microscopic HFB PECs
onto the IBM-2 parameter space.
Figure~\ref{fig:pec}
compares the predicted IBM-2 PECs
with those of the microscopic
Gogny-D1S HFB PECs for Sm nuclei.
One can see that
the PGNN reproduces to a certain
degree of accuracy the overall features
of the HFB PECs, such as the
curvature near the energy minimum
and the absolute energy depth
of the HFB potentials, for
each isotope under investigation.
Furthermore, the rapid deepening
of the potential wells and the shift
of the location of the
energy minimum, $\beta_{\rm min}$,
toward a larger deformation for
$N \geqslant 90$ in the HFB calculations
are reasonably described.
This demonstrates that the PGNN does
not merely perform a reasonable
numerical fit, but incorporates
the information about the nuclear
structural evolution encoded
in the Gogny-D1S force.

\begin{figure}[t]
\centering
\includegraphics[width=.9\linewidth]{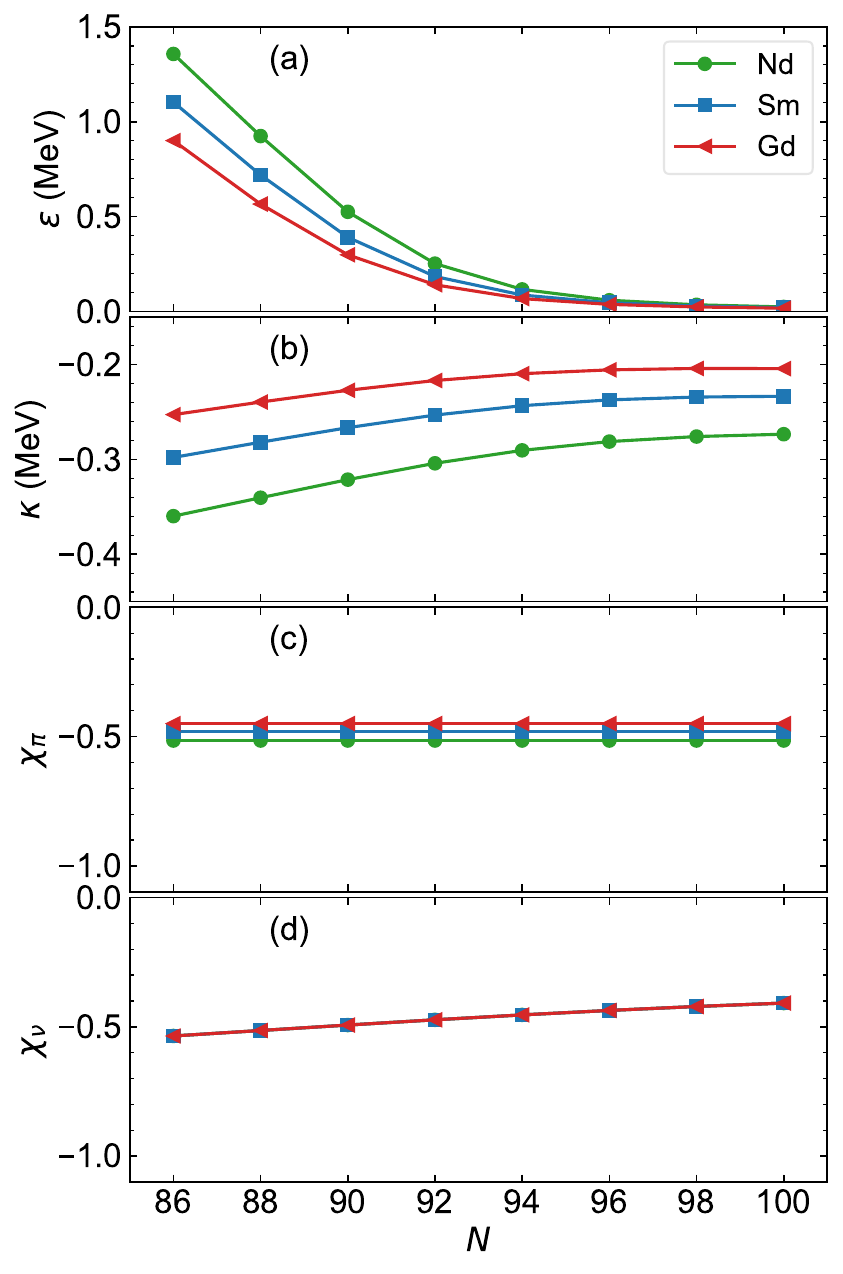}
\caption{
Derived IBM-2 parameters
for Nd, Sm, and Gd
isotopes as functions of
the neutron number $N$: (a) Single-$d$-boson
energy, $\epsilon$, (b) quadrupole-quadrupole
interaction strength, $\kappa$,
(c) proton and (d) neutron
shape parameters, $\chi_{\pi}$
and $\chi_{\nu}$, respectively.}
\label{fig:para}
\end{figure}

The robustness of the PGNN framework
is most clearly demonstrated
in the systematic evolution of
the extracted IBM-2 parameters,
$\{\epsilon, \kappa, \chi_\pi, \chi_\nu\}$.
One indeed observes in Fig.~\ref{fig:para}
a high degree of
similarity in the $N$-dependence
of the predicted parameters
between the three isotopic chains.
As $N$ increases,
the derived $\epsilon$ and $\kappa$
gradually decrease in magnitude
toward the middle of the neutron major shell,
while $\chi_{\pi}$ remains constant.
These results are consistent with
the conventional IBM-2 PES mapping
\cite{nomura2008} and OAI
mapping \cite{OAI}.
What is interesting is that,
while the present $\chi_{\nu}$ values
obtained by the PGNN
do not exhibit strong $N$-dependence
and gradually decrease in magnitude with $N$,
in the above earlier IBM-2 mapping
$\chi_{\nu}$ was shown to increase
in magnitude significantly with $N$
to be closer to the SU(3) limit,
$\chi_{\nu}=-\sqrt{7}/2$ \cite{IBM},
in the deformed region.
The reason for this deviation to
occur could be that,
since the shape of the IBM-2 PEC
is rather sensitive to $\epsilon$
and $\kappa$, the NN tries to reproduce
the topology of the HFB PEC
near the energy minimum,
in particular, the location of $\beta_{\rm min}$,
by reducing $\epsilon$ and
$|\kappa|$ rather than
increasing $|\chi_{\nu}|$.
Another reason may be that our analysis
is restricted to axial symmetry.
Inclusion of the triaxial degree
of freedom may affect the derived
values and $N$-dependence of $\chi_{\nu}$.

Energy levels are calculated
for each nucleus
by numerically diagonalizing
\cite{NPBOS}
the mapped boson Hamiltonian,
$\hat H_{\rm IBM}$ \eqref{eq:hb},
with the set of the derived
parameters shown in Fig.~\ref{fig:para}.
Generally, the IBM, when it is formulated
microscopically, is known to
overestimate significantly
the ground-state-band
energy levels in deformed nuclei
\cite{otsuka1979phd,otsuka1981,
nomura2008,nomura2011rot},
even though their
overall systematic behaviors with $N$
are consistent with experiment.
This discrepancy turns out to
arise also in the present PGNN-IBM
calculations, which overestimate
the observed energy levels
in deformed nuclei with $N\geqslant90$
by a factor of $\approx1.5-1.8$.
The overestimates could
be attributed \cite{nomura2011rot}
to the fact that a certain rotational
response of the nucleonic system
cannot be taken into account in the
boson system through the
Hamiltonian of \eqref{eq:hb}.
To resolve the problem,
by employing the prescription of
Ref.~\cite{nomura2011rot},
we add to the IBM-2 Hamiltonian
the term representing the rotational
correction, $\hat{L} \cdot \hat{L}$,
as
\begin{equation}
\label{eq:LL}
 \hat H_{\rm IBM}
\rightarrow
\hat H_{\rm IBM} + \kappa' \hat{L} \cdot \hat{L}
\; ,
\end{equation}
where $\hat L = \hat L_{\pi} + \hat L_{\nu}$
with
$\hat L_{\rho}=\sqrt{10}[d^{\+}_{\rho}\tilde d_{\rho}]^{(1)}$
being the boson angular momentum operator.
The $\hat L \cdot \hat L$
strength parameter, $\kappa'$,
is uniquely determined by equating
the bosonic cranking
moment of inertia \cite{schaaser1986}
calculated at the energy
minimum of the IBM-2 PEC
to the corresponding quantity
\cite{CEA}
computed by the Gogny-HFB method
using the Thouless-Valatin formula \cite{TV}
at the minimum of the HFB PEC.
Since this process is performed
independently of the PEC mapping,
the other parameters
that are specified by the NN
remains unaffected.
The derived $\kappa'$ values
are approximately
equal to $-0.01$ MeV for the Sm isotopes.

\begin{figure}[t]
\includegraphics[width=.9\linewidth]{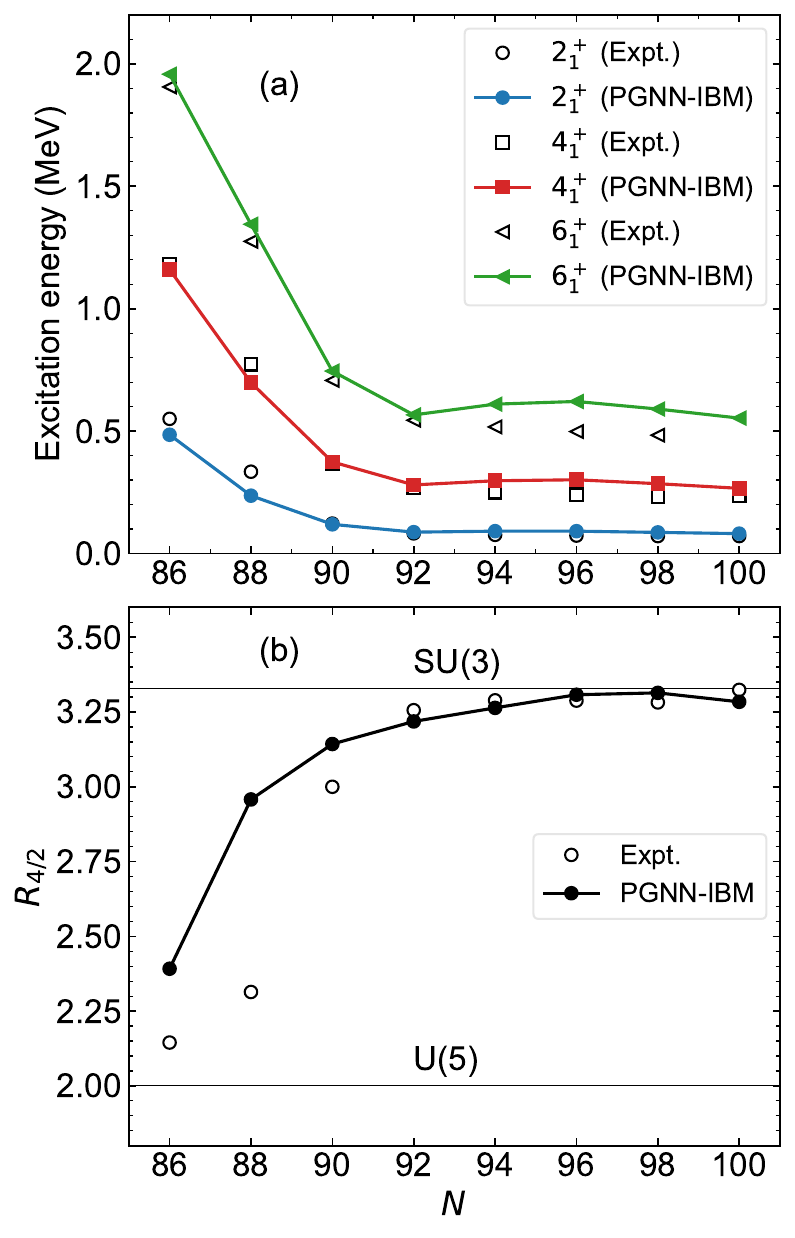}
\caption{Calculated and experimental
(a) excitation energies of the yrast
states, $2^+_1$, $4^+_1$,
and $6^+_1$, and
(b) $R_{4/2}=E(4^+_1)/E(2^+_1)$ ratios
in Sm isotopes.
The $R_{4/2}$ values in the
U(5) and SU(3) limits \cite{IBM}, 2 and 10/3,
respectively, are indicated in (b).
The experimental data are adopted
from the NNDC database \cite{NNDC}.}
\label{fig:level}
\end{figure}

The calculated ground-state-band levels
in the PGNN-IBM, incorporating
the rotational correction
through \eqref{eq:LL}, are
shown in Fig.~\ref{fig:level}(a).
One can see that
the experimental energy levels
and their decreases as functions of $N$
are well reproduced by the PGNN-IBM.
This reflects that
the model is able to capture
the onset of collectivity.
As a signature of the QPT,
Fig.~\ref{fig:level}(b) compares
the experimental \cite{NNDC}
and theoretical $R_{4/2}$ ratios.
The PGNN-IBM qualitatively
captures the global trend of the
structural evolution, that is,
the transition from the vibrational U(5)
limit (with $R_{4/2}=2.0$)
to the rotational SU(3) limit ($R_{4/2}=10/3$)
\cite{IBM}.
The predicted $R_{4/2}$ values
exhibit an earlier rise than the
experimental ones, specifically,
from $N=86$ to 88.
This indicates that the PGNN-derived
Hamiltonian predicts the onset of
deformation at slightly lower neutron numbers
than observed experimentally.
This discrepancy is primarily attributed
to the characteristics of the
underlying HFB energy landscapes:
The Gogny-D1S HFB PECs generally
exhibit the onset of deformation
at a lower neutron number than
expected from experiment;
for $N=86$ and $88$,
while the observed $R_{4/2}$
ratios still indicate a vibrational character,
the HFB calculations already show
a nonzero $\beta_{\rm min}$
(cf. Fig.~\ref{fig:pec}).

Concerning the properties of
states in nonyrast bands, e.g., $\beta$
and $\gamma$ vibrational bands,
our PGNN-IBM approach predicts their
bandhead energies
to be much higher than the observed ones.
This problem is generally encountered
in the DFT-mapped IBM-2
calculations, and is attributed to
some missing correlations such as
configuration mixing,
which was suggested to be crucial
for interpreting
structure of rare-earth nuclei
\cite{garrett2022,tsunoda2023,leoni2024,nomura2025bb-2}.
An accurate description of the nonyrast
states is an interesting future
work, but is also beyond the scope of the
present work.

To conclude, we have developed
a neural network to establish
a microscopic derivation of the IBM-2
Hamiltonian parameters from the
Gogny-HFB PECs.
Our hybrid approach successfully
bridges the gap between the
microscopic and algebraic models
without any manual parameter tuning.
In a systematic analysis of
the Nd, Sm, and Gd isotopic chains,
we developed a mapping
architecture characterized by its
physics-informed branching structure.
It demonstrated consistency
with earlier microscopic IBM-2
studies by enforcing the independence
of proton and neutron shape parameters,
and successfully captured the
essential signatures of the shape QPT.
As a future work,
we plan to extend
the training data to include
triaxial PESs with
both the $\beta$ and $\gamma$
degrees of freedom.
To process these higher-dimensional
energy landscapes,
convolutional NN architectures
equipped with a differentiable
physics layer will be introduced.
Such an extension of the PGNN
is a pathway to describe collective
excitations associated with
more general nuclear shapes,
providing a robust microscopic
foundation of the IBM in the entire
nuclear chart.
In addition, the proposed NN
could be applied to predict bosonic
energy surfaces and Hamiltonian parameters,
particularly, for those
nuclei in the mass region in which
nuclear structure changes
rapidly from one nucleus to another,
and for which the direct spectroscopic
predictions within the DFT
is challenging.
This presents a certain step
toward a consistent theoretical
description of the
nuclear deformations
with machine learning.

\section*{Acknowledgements}

This work has been supported
by JSPS KAKENHI Grant No. JP25K07293.

\bibliographystyle{elsarticle-num} 
\bibliography{refs}

\end{document}